\documentclass[conference]{IEEEtran}
\IEEEoverridecommandlockouts
\usepackage{cite}
\usepackage{amsmath,amssymb,amsfonts}
\usepackage{algorithmic}
\usepackage{graphicx}
\usepackage{textcomp}
\usepackage{xcolor}
\usepackage{balance}
\usepackage[hidelinks]{hyperref}
\usepackage{caption}
\usepackage{subcaption}
\usepackage{array}
\def\BibTeX{{\rm B\kern-.05em{\sc i\kern-.025em b}\kern-.08em
    T\kern-.1667em\lower.7ex\hbox{E}\kern-.125emX}}
\begin{document}
\newcommand\note[2]{\color{#1}\bf #2}
\newcommand\pb[1]{{\note{red}{paolo: #1}}}
\newcommand\af[1]{{\note{blue}{angelo: #1}}}
\newcommand\ac[1]{{\note{orange}{alessandro: #1}}}
\newcommand\ab[1]{{\note{green}{armir: #1}}}

% To reduce padding between image and its caption
\captionsetup[figure]{skip=4pt}

%%%%
% Acronym definition 
%%%%
\newcommand\mec{{MEC}}
\newcommand\mecapp{{MEC-App}}
\newcommand\orchestrator{{MEC-O}}
\newcommand\ualcmp{{UALCMP}}
\newcommand\mechost{{MEC-H}}
\newcommand\vi{{VI}}
\newcommand\vim{{VIM}}
\newcommand\mecplatform{{MEC-P}}
\newcommand\mecpm{{MEC-PM}}
\newcommand\locations{{LC}}
\newcommand\rni{{RNI}}
\newcommand\ned{{NED}}
\newcommand\omnetpp{{OMNeT++}}
\newcommand\server{{Broker}}
\newcommand\aoi{{AoI}}

% 5G e' necessario in quanto tutto il system desing e' stato pensato in un applicazione di MEC nel 5G \cite{kekki2018mec}
\title{A Novel Design for Advanced 5G Deployment Environments with Virtualized Resources at Vehicular and MEC Nodes}%*\\
%An ETSI MEC compliant extension for resources at the far edge
%Extending the ETSI MEC reach to vehicular resources at the edge
%A simulation model to exploit complementarities of ETSI MEC and Far edge vehicular resources
%Extending ETSI MEC to reach vehicular resources at the far-edge of 5G networks

\author{\IEEEauthorblockN{Angelo Feraudo, Alessandro Calvio, Armir Bujari, Paolo Bellavista}
\IEEEauthorblockA{\textit{Department of Computer Science and Engineering} \\
\textit{University of Bologna}\\
Bologna, Italy \\
name.surname@unibo.it}
% \and
% \IEEEauthorblockN{Alessandro Calvio}
% \and
% \IEEEauthorblockN{Armir Bujari}
% \and
% \IEEEauthorblockN{Paolo Bellavista}
}

\maketitle

\begin{abstract}
%--> importanza MEC
%--> extensione: voglio andare a raggiungere i veicoli 
%--> importanza di modello simulazione
%--> validazione e assessment del modello in a novel scenario (fed. learning trasferimento stato o altro)
IoT and edge computing are profoundly changing the information era, bringing a hyper-connected and context-aware computing environment to reality. Connected vehicles are a critical outcome of this synergy, allowing for the seamless interconnection of autonomous mobile/fixed objects, giving rise to a decentralized vehicle-to-everything (V2X) paradigm. On this front, the European Telecommunications Standards Institute (ETSI) proposed the Multi-Access Edge Computing (MEC) standard, addressing the execution of cloud-like services at the very edge of the infrastructure, thus facilitating the support of low-latency services at the far-edge. In this article, we go a step further and propose a novel ETSI MEC-compliant architecture that fully exploits the synergies between the edge and far-edge, extending the pool of virtualized resources available at MEC nodes with vehicular ones found in the vicinity. In particular, our approach allows vehicle entities to access and partake in a negotiation process embodying a rewarding scheme, while addressing resource volatility as vehicles join and leave the resource pool. To demonstrate the viability and flexibility of our proposed approach, we have built an ETSI MEC-compliant simulation model, which could be tailored to distribute application requests based on the availability of both local and remote resources, managing their transparent migration and execution. In addition, the paper reports on the experimental validation of our proposal in a 5G network setting, contrasting different service delivery modes, by highlighting the potential of the dynamic exploitation of far-edge vehicular resources.
\end{abstract}

\begin{IEEEkeywords}
Multi-access Edge Computing, MEC, Vehicular Computing, VANET, 5G
\end{IEEEkeywords}

\section{Introduction}

Thanks to the vast improvements in computing technologies and the pervasive deployment of next-generation communication networks, it is estimated that every new vehicle will be connected in the near future, embodying the potential of a fully-fledged mobile computing platform where vehicles serve as computation nodes for a diverse range of services. Different from vehicular networking~\cite{7513432}, which serves as a communication enabler for applications associated with transportation, vehicle computing focuses on the computation function and emphasizes the promising role it embodies towards the implementation of a pervasive context-aware computing environment. This computing environment can play a major role in future ICT systems for supporting applications like Intelligent Transportation Systems (ITS) and full-scale smart cities~\cite{WHAIDUZZAMAN2014325}.

Mobility-as-a-service and high-definition (HD) map generation are examples of such services, provisioned spanning cloud-to-vehicle resources where computationally heavy tasks are offloaded to resource-hungry cloud-based nodes. However, as the number of embedded sensors and smart vehicles grows, the amount of in-vehicle generated data may represent a serious problem for the infrastructure~\cite{aecc}. At the same time, cloud-based solutions are generally unfit to serve delay-sensitive application scenarios, which are currently served by shifting computation at the edge of the network while providing services and data preprocessing functions.

On this front, the European Telecommunications Standards Institute (ETSI) has worked on the standardization of a cloud platform co-located at the edge of the network, including the Radio Access Network (RAN), named Multi-Access Edge Computing (\mec)~\cite{etsiwebsiterefarch}. The standardized architecture includes functional components tasked with the management and orchestration of edge resources, managing \mec~applications' life-cycle, and providing standardized reference points to access the services. ETSI \mec{} augments legacy radio units with cloud-like computing capabilities, thus representing a pillar technology for (beyond) 5G cellular networks, allowing application deployment spanning cloud/edge resources~\cite{kekki2018mec}.  

New opportunities and challenges arise as a growing number of businesses start to exploit the shared edge-cloud environment. In contrast to traditional cloud deployment environments in data centers, the edge has limited resources and may not always be able to satisfy application demands for resources and associated QoS. Moreover, the technical challenges associated with advanced edge infrastructures are exacerbated by the convergence trends meant to make sure an end-user can access the whole range of subscribed services. % whatever device technology, wherever the user is connected to the network, and whether the user is in motion or not. %Recently, several research efforts highlighted as present-day vehicles include powerful resources that can be deployed as small data center~\cite{olariu2011taking, 7018198, 8936985,7415983}.
% For this reason, the need to identify new resources that can support the infrastructure dynamically when overloads happen has arisen. Today vehicles are equipped with a rich set of computing, data storage, communication, and sensor resources in their onboard computer unit. Based on current vehicular networking standards, cellular vehicle-to-everything (C-V2X) communication enables vehicles to wirelessly connect and cooperate with each other and with their surroundings, including road infrastructure, pedestrians, and so on. Thus, it is believed that vehicles will play a major role in future Information and Communication Systems (ICT) systems for supporting applications like Intelligent Transportation Systems (ITS) and full-scale smart cities. Because of these characteristics, vehicles can be considered true computational nodes capable of supporting the computing infrastructure.
For these reasons, the need has emerged to identify new resources that can support the edge infrastructure dynamically, thus enabling service availability in dense and congested deployment scenarios. Toward this end, some examples of applications that may benefit from such a dynamic scenario involve decentralized learning contexts, where opportunistic resources can improve the overall learning process in terms of data quality and training/prediction speed. Additionally, in a smart city context, the availability of more resources could enable the creation of edge-enabled digital twins to keep track of assets present in the city and execute computationally demanding simulation tasks.

Taking a step towards the implementation of the aforementioned pervasive computing environment, our proposal extends the MEC resource pool so to leverage far-edge device resources, exposed and made available in a standardized way. To this end, we propose a novel ETSI \mec-compliant architecture that can tap into far-edge resources within an Area of Interest (\aoi). Node resources are registered at the edge resource pool, exposed and accessed via standardized interfaces. To handle node mobility, the architecture assists the application migration by triggering the user context transfer to \mec~applications running on nodes leaving the \aoi. To demonstrate the viability of our proposal, we have developed a simulation model, readily available for researchers in \cite{githubrepo}, that extends the \mec~infrastructure with vehicular onboard resources. Our simulation model supports \mec~application deployment on vehicle resources and manages their volatility as vehicles enter and leave the AoI. We evaluate the proposed approach by benefiting from two real-world datasets (about user mobility and parking lot occupancy) where there is the need to manage service migrations triggered by vehicle mobility. Finally, let us highlight that this work represents the first effort at enabling the deployment of vehicular resources in a multi-vendor and multi-operator scenario, through the exploitation of (the original extension of) an established ETSI standard.

% To demonstrate the viability of our proposal in 5G scenarios, we developed a simulation model that extends the \mec~infrastructure through parked vehicles onboard resources. The model supports \mec~application deployment on vehicle resources, and manages their volatility as vehicles enter and leave the parking area. In this sense, the work analyzed two real-world datasets to provide an in-depth study of service migrations generated by vehicle mobility. Moreover, this work represents the first effort at enabling the deployment of vehicular resources in a multi-vendor and multi-operator scenario, as it relies on an extension of an established standard. 

%Additionally, this solution is one of the few proposals that perform app migration and the only one that makes use of standard APIs. 
% \af{Add the fact that we are providing an analysis}
% The paper provides extensive simulations to demonstrate the feasibility and the performance of our framework in 5G contexts. Moreover, we analyzed two real-world datasets to provide an in-depth study of cars and User Equipment (UE) behaviors. This study allowed us to build the first environment tracing service migration generated by cars mobility and to test system performance during migration.

\begin{figure*}[h]
  \centering
  \includegraphics[width=\textwidth]{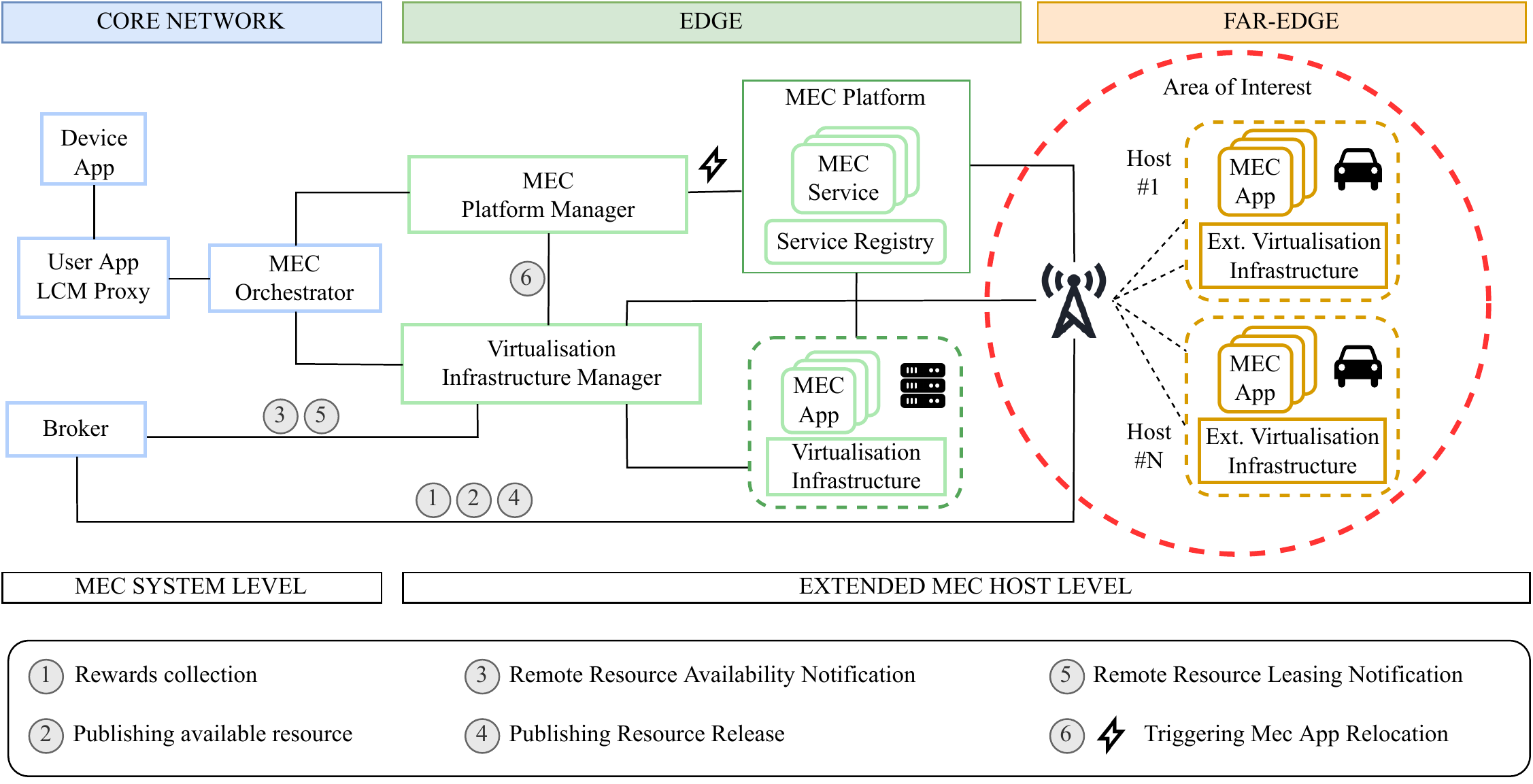}
  \caption{\mec{} extension architecture diagram}
  \label{fig:extension_architecture}
    \vspace{-0.4cm}
\end{figure*}
\section{Background}\label{sec:bg}
% \label{section:backround}

% IMPORTANT NOTE:
% Here we refer to the standard architecture fig 6-1 {https://www.etsi.org/deliver/etsi_gs/MEC/001_099/003/03.01.01_60/gs_MEC003v030101p.pdf}, which does not introduce the concept of virtual network function as in its variant (section 6.2 linked document)
This section provides a brief description of the main functional elements of the ETSI \mec{} architecture used as the standard reference basis in our design, along with details about the simulation tools adopted to develop the extended model.

\subsection{ETSI MEC}
The \mec{} standard by ETSI proposes an architecture \cite{etsiwebsiterefarch} introducing cloud computing
capabilities at the very edge of the network, offering the ability to run so-called MEC Applications (\mecapp{}) within a virtualized and multi-tenant environment. The \mec{} architecture comprises two main parts: the system and host levels.
The system level represents the entry point for users to request service execution and it is usually deployed in the core network. The central element, the \mec{} Orchestrator (\orchestrator{}), has a view of the \mec{} Hosts (\mechost s) present in a particular area and is responsible for choosing the best one according to the requirements of the requested service, triggering both the instantiation and termination of applications. In this context, the User Application LifeCycle Management Proxy (\ualcmp{}) acts as an intermediary for the users by supporting instantiation and termination requests forwarded to the \orchestrator{}.

The host level lies at the edge of the network and realizes and manages the virtualization platform where applications are deployed. At this tier of the network, the \mechost{} is the main entity providing computational, network, and storage resources for the \mecapp{}s by exploiting the \mec{} Platform (\mecplatform{}) and the Virtualization Infrastructure (\vi{}). The \mecplatform{} exposes a service registry in which applications can discover, offer and consume standard \mec{} services defined by ETSI; at the moment of writing, the standard supports four types of services that each \mechost{} might maintain: Radio Network Information Service, Location Service, Traffic Management Service, and  Application Mobility Service (AMS)~\cite{etsiamsapi} conceived to handle \mecapp{} migrations between edge nodes. In this context, each application runs as a virtual machine or container 
on top of the \vi, and it optionally interacts with the \mecplatform{} and its registry to take advantage of the standard \mec{} services.
Finally, alongside the \mechost{}, the Virtualization Infrastructure Manager (\vim{}) and the \mec{} Platform Manager (\mecpm{}) serve as access points for the \orchestrator{} to the lower layer. The \vim{} has the task of managing and releasing the virtualized resources and appropriately configuring the \vi{} to run software images, while the \mecpm{} handles all the components functions related to a specific \mechost{}.

\begin{table}
\centering
\caption{Table of acronyms for MEC elements}
\resizebox{\linewidth}{!}{{\renewcommand{\arraystretch}{1.3}%Vertical padding factor
\begin{tabular}{c c}
\hline
 \textbf{Abbreviation} & \textbf{Definition} \\
 \hline
 AMS & Application Mobility Service \\
 % \hline
 MEC & Multi-access Edge Computing \\
 % \hline
 MEC-App & MEC Application \\
 % \hline
 MEC-H & MEC Host \\
 % \hline
 MEC-O & MEC Orchestrator \\
 % \hline
 MEC-P & MEC Platform \\
 % \hline
 MEC-PM & MEC Platform Manager \\
 % \hline
 UALCMP & User Application LifeCycle Management Proxy \\
 % \hline
 VI & Virtualisation Infrastructure \\
 % \hline
 VIM & Virtualisation Infrastructure Manager \\
 \hline
\end{tabular}}}
\label{table:mec_acr}
\end{table}

\subsection{\omnetpp{} and Simu5G}\label{subsec:bgsimtool}
\omnetpp{}\cite{omnetwebsite} is a discrete event simulator framework used to model and build general-purpose simulations. It proposes a modular architecture based on components that can be arranged to create simulation models easily and effectively. 
%Armir: Thanks to this feature, \omnetpp{} has allowed it to gain popularity within the scientific community as a network simulation tool and, to date, it is supported by several third-party projects that continue to extend it with new models.

%It proposes a programming system based on parameterized components called modules that, if properly designed, can be reused easily and effectively. The modules are divided into simple and compound and are arranged, with a declarative approach, in a hierarchical structure that represents the final model of the simulation. Moreover, the communication between components relies on message exchange that leverages a system of gates used to define the input and output entry points for modules. The highly modular architecture of \omnetpp{} has allowed it to gain popularity within the scientific community as a network simulation tool and, to date, it is supported by several third-party projects that continue to extend it with new models.

Build on top of \omnetpp{} is Simu5G~\cite{simu5g_1}, a simulation library containing a collection of models and components useful for creating arbitrarily complex end-to-end scenarios involving 5G radio networks. Simu5G models both the core network and the RAN of a 5G network through the implementation of 3GPP-compliant protocols and a physical transmission system based on a set of customizable channels. The simulations created with this library can also leverage heterogeneous models related to gNBs base stations 
% that communicate via X2 interface 
while supporting handover and inter-cell interference coordination. Indeed, Simu5G can be used to analyze transition scenarios, from 4G to 5G networks, thanks to the ability to use dual connectivity (X2) between eNB and gNB access types. Among the various implemented models, Simu5G provides an implementation of the ETSI MEC standard with its main components (i.e., \orchestrator, \mechost, etc.) and the ability to create applications that communicate via ETSI-compliant interfaces with other elements of the MEC ecosystem. Running applications can either be self-contained or take advantage of the presence of standard MEC services; currently, the framework implements the Location Service and the Radio Network Interface Service. A significant feature of the tool is that Simu5G can also be employed as a real-time emulator to replace simulation elements 
%(5G network, MEC elements, etc.)
with real devices and thus use the same code for simulation and prototyping. Additionally, this also allows for more realistic and reliable data collection. 

Our simulation model works within the \omnetpp{} simulation tool and relies on the Simu5G communication library. It provides a new interpretation of the ETSI \mec{} functional elements enabling dynamic resource acquisition at \mec{} Host level. 

While preserving the general aspect of our study and without loss of generality, in the following section, we survey prior research effort on key design features our solution embodies.

%Armir: descrivere in modo sintetico il nostro contributo - costruito sopra Simu5G - e anticipare la sec. successiva.
% {\color{red}@Angelo: c'era un comento mio anche prima. 1 frase di delta w.r.t Simu, e.g., broker}

\section{Related Work}
Considerable research effort has been devoted in the past, advocating for the use of vehicular resources to improve service delivery at the edge~\cite{8936985}. %and/or the support for alternate service delivery models without strict infrastructure reliance~\cite{8936985}. 
The concept of Vehicular Cloud Computing was initially proposed in~\cite{abuelela2010} and~\cite{6257116}. Abuelela and Olariu~\cite{abuelela2010} introduced the concept of Vehicular Cloud (VC), where vehicular resources are exploited as a mean to provide diverse community services. Similarly, Gerla in~\cite{6257116} outlined two applications of VC in which vehicles not only act as datacentres but also as observers of the environment. The service delivery model discussed in these works embodies new challenges when compared to the traditional, infrastructure-based one, such as task scheduling and distribution, resource volatility and acquisition, to be attributed to the unpredictability of the environment, i.e., node mobility. 

%Armir: To deal with the volatility of topological information, Arif et al. in~\cite{6143927} modeled the residency time of vehicles in the parking lot of an airport to create an accurate picture of the number of available resources.
Counteracting the mobility phenomenon, the authors in~\cite{5935198, 9475490} proposed to use parked vehicles as relay nodes, which can help improve connectivity and augment the chances of message delivery. 
Conversely, other works focused on exploiting vehicle resources for edge application execution~\cite{8522034, 9344808, 9366768, 9709120}. On this front, Huang \emph{et al}.\cite{8522034} proposed a centralized architecture, where the central node (i.e., nodes at the edge of the network) receive task request which are then distributed as sub-tasks on selected parked vehicles.
Similar in its objective, the work in~\cite{9344808} proposes a decentralized approach, offloading task execution to vehicle resources available nearby. Addressing a practical challenge to service delivery, Li \emph{et al}.~\cite{8463481} defined a contract-based incentive mechanism to persuade vehicle owners to rent out their resources. Similarly, the authors in~\cite{9475490} propose an auction-based model where participating nodes compete to lend their resources in an extended vehicular resource pool. %to use vehicle resources as network traffic forwarding. 

%Due to mobility, the vehicular resource pool might be subject to continuous changes in terms of capacity, and in task offloading-based algorithm it is a necessary system primitive which needs to be carefully considered.
Due to mobility, the vehicular resource pool might be subject to continuous changes in terms of capacity and task offloading decisions require carefully consideration. The necessity for this might also arise due to an inaccurate estimate of residual resource availability. In this context, most of the works propose algorithmic strategies used to evaluate the probability of nodes to complete task execution~\cite{8522034, 9123902, 9344808, 9366768, 9709120}. However, these approaches are probabilistic, neglecting also practical considerations such as nodes refusing to partake in the resource pool. In this direction, the authors in ~\cite{ge2020two} present a two-stage algorithm that handles service migration from one service provider to another in a vehicular network context. The algorithm relies on several metrics, such as average latency and energy consumption, to properly select the next service provider.
%Armir: dire 1/2 frasi ++ sul lavoro sopra, e.g., come fanno la evaluation. Inoltre rendere chiaro se l'offloading a solo infrastructure->VANET o cosa...
Although there has been some research effort addressing task scheduling and resource allocation problems in the vehicular network tier, these proposals neglect and make no consideration of the challenges arising in multi-vendor and multi-domain environments. Furthermore, the migration problem does not seem to concern many authors. 

In this paper, we propose an ETSI \mec-compliant architecture and an accompanying simulation model, extending the edge resource pool to contemplate far-edge (vehicular) resource infrastructures. Vehicular resources can be transparently accessed and made readily available through standardized interfaces. Our proposal has built-in mechanisms capable of deploying and distributing applications on the available resource pool, while addressing resource volatility issues (i.e., nodes joining/leaving) via a transparent migration mechanism that exploits already available constructs. 
As an extension of the ETSI \mec~standard, the model allows us to deal with some of the aforementioned challenges, providing better integration with cloud resources and enabling coexistence among heterogeneous technologies.

\section{Our Proposed MEC Extension for the Dynamic Exploitation of Neighbor Vehicular Resources}\label{sec:sysdes}

%Armir(0) broker centralizato vs. distribuito (cloud-edge)
Our work stems from the observation that \mechost{}s (i.e. edge nodes) have limited capabilities when compared to cloud-backed ones. %and it is the entity that provides resources to run \mecapp{}s on top of the \vi{}. 
The proposal turns the \mechost{} into a logical entity that can leverage multiple \vi{}s, dynamically adding and removing localized computational resources.

% VIM exaplanation
Referring to Fig.~\ref{fig:extension_architecture}, our proposal allows the inclusion of the far-edge (vehicular) layer in cloud continuum deployment environments. Thus, in addition to locally defined hosts (edge nodes), it involves also remote host resources that are reachable and added dynamically via the RAN (e.g., 5G RAN). The approach entails some changes in the \mec{} traditional architecture in terms of structure and interactions (steps \textcircled{1}-\textcircled{6} in Fig.~\ref{fig:extension_architecture}). The main component to be affected is the \vim{} that is in charge of administering the host resources. In our design, it handles a heterogeneous pool of distributed resources and is aware of the single contributions that each host brings in terms of capacity. In this context, it is desirable to differentiate between infrastructure resources and the more transient ones, dynamically joining the \mechost{} thanks to vehicle availability in the neighborhood.

% Protocol description - external entity
%Armir (1): service discovery | network or vehicule-initiated (assisted).
% FLOW OF THE SECTION
%   1) Description of the main element: a mechanism to handle the join and leave of cars
%   2) How we designed it? Introduction of AoI as high-level abstraction - we use base station converage
%   3) Introduction of a new entity: the broker
%   4) How the broker is involved? Car discovery vs vehicle initiated - explain
%   5) what we decided (?)
%   6) Description of the joining protocol
%%%%
% writing...
%%%%
% Broker description - a mechanism to handle resources joining and leaving
Our architecture proposal implies the creation of a mechanism to handle far-edge resources joining and leaving the \mechost{} resource pool. Thus, each \mechost{} defines an \aoi{} within which far-edge nodes (hosts) can decide whether to provide their onboard resources. To model this mechanism, we decided to involve a new external component in the core network, named \server{}, which handles the resource pooling of several \mechost{}s. To this end, the \server{} relies on a publish-subscribe model to collect \mechost{} \aoi{} subscriptions and manage notifications whenever new devices enter the area. The \aoi{} may depend on where \mechost{}s are located, i.e., at the network edge or network aggregation points~\cite{kekki2018mec}. 
%Armir: what does decoupled model mean? - pub/sub model
% However, thanks to the decoupled model employed, our architecture supports the opportunity to dynamically aggregate far-edge resources and create \mec{} compliant node in any area within network coverage.
% PROCEDURE DESCRIPTION:
% MEC-HOST SUB
% RESOURCE SUB
% RESOURCE RELEASE
% MEC-host subscription
Hence, each \mechost{}, at bootstrap time, subscribes to an \aoi{} that might coincide with one or more zones, which typically correspond to the coverage of the associated gNBs.

%The joining procedure may be initiated in two ways: \server{} asking each entering device its available resources by encouraging them throug; devices entering the \aoi{}, notifying its presence, and asks the \server{} for a potential reward.

% Joining of new resources
% Finally, a device entering an \aoi{} and deciding to provide its resources triggers the notification chain that allows increasing \mechost{} computational capacity.
A reward system encourages far-edge nodes to lease their computational capacity and join the resource pool. Two different schemes of the procedure are envisioned: network- or vehicle-initiated. The former requires the \server{} to provide each far-edge node in the \aoi{} a set of rewards to incentivise resource leasing. The latter (steps \textcircled{1}-\textcircled{3} ) expects far-edge nodes to manifest their intention to join the pool, by asking for available rewards contextualized to the \aoi{}. In both cases, if the device finds acceptable terms, it can confirm the intent to participate by communicating to the \server{} the set of leased resources. The current design adopts the second approach, whereby vehicles obtain available rewards and can autonomously decide whether completing resource registration. Similarly, when one of them leaves the \aoi{}, it notifies the \server{}, which forwards the release request to the \vim{} managing that area. The latter in turn removes the concerned resources from those available in the pool. 

% Resource release
The resource release procedure (steps \textcircled{4}-\textcircled{6}) requires more attention as the departing host may have applications running on it. In such a case, after receiving the release notification, the \vim{} triggers the migration procedure (step \textcircled{6}) to move running apps from one host to another and thus maintaining service continuity with very low service interruptions. Both registered hosts in the \aoi{} and the local infrastructure of the \mechost{} are eligible to support the migration operation and embrace the new application(s).
The AMS, defined in the ETSI standard, currently supports app migration in environments encompassing multiple edge nodes. %This service is designed to support app mobility in environments with multiple edge nodes. 
Our extension has been designed to work in this perspective by extending the service to further support intra-host migration in a standard way. Furthermore, 
%in such extension, 
it is \mecpm{} that is identified as the main component that, during the procedure, acts as an intermediary node between the AMS and the \vim{} for new app allocations.
% A device (host) that enters the \aoi{} can request available rewards in order to decide whether to join the pool or not. 

%can decide whether to join the pool or not based on a reward system, which encourages the leasing of resources for devices that manifest their intention to join the pool.
%Armir(Opt.): se si da anche un idea vaga delle componenti coinvolte e' meglio..

%Armir(Opt): In the current implementation, we provide a basic monetary reward which is always accepted by the party, leasing in return all available resources. As a future extension, we envision to generalize the approach by introducing a mechanism implementing pluggable reward schemes.  - done
%A pub/sub messaging system is used to forward each request to the correct \vim{} and complete the resource registration.
% Similarly, when a far-edge device to leave the \aoi{}, it notifies the \server{}, which then forwards the release request to the \vim{} managing that area. 
%Armir(2): IMPORTANTE DIRE CHE FAR-EDGE RESOURCES ARE SHARED. OTHER APPS MIGHT PREEMPT THE EXECUTION OF OTHERS ---> HENCE MIGRATION NEEDED FOR THE FORMER APP.

Finally, concerning the resource allocation and scheduling approach, the \vim{}s initial selection results in a set of hosts eligible for app deployment; the scheduling phase, leading to the identification of a single host, can be done by ordering nodes depending on strategies that might favor certain aspects over others. Some metric examples are the average latency time between a host and the central infrastructure or the probability of a node to further contribute to the resource pool (i.e., based on historical data).

The proposed design approach paves the way for innovative application scenarios, which go beyond the state-of-the-art far-edge computing ones targeted nowadays, such as task offloading and content caching. For example, our proposal could be a key enabling element for the hosting of Federated Learning \cite{MLSYS2019} enabled environments at the far-edge layer. Specifically, a \mecapp{} (federated server) running on \mechost{} local resource infrastructure can coordinate other \mecapp s (federated clients) deployed on remote resources during the training on local data. The former can choose federated clients by using the model descriptors and information collected through the \mec{} standard API, while the latter, after receiving a federated model, can start the training procedure by relying on their status, local data, and received rewards.
%Just a note: our approach neglects security issues on how malicious nodes can affect mechost operational aspects}%
%Our extension moves in that direction, extending the service to also support intra-host migration in a standard way. \af{may be this first part can be cut} While in the general case it is the \orchestrator{} that handles the messages exchanged during the procedure, in our scenario the \mecpm{}  acts as an intermediary between the AMS and the VIM for new app allocations.%

%In the next section, we validated the proposed architecture by developing a simulation model that relies on parked vehicle resources to distribute \mec{} applications and service requests.
% The next section shows the validation of the architecture proposedby creating in a simulated environment and performance evaluation related to a concrete use case.
 \begin{figure}[b]
    \centering
    \includegraphics[width=\linewidth]{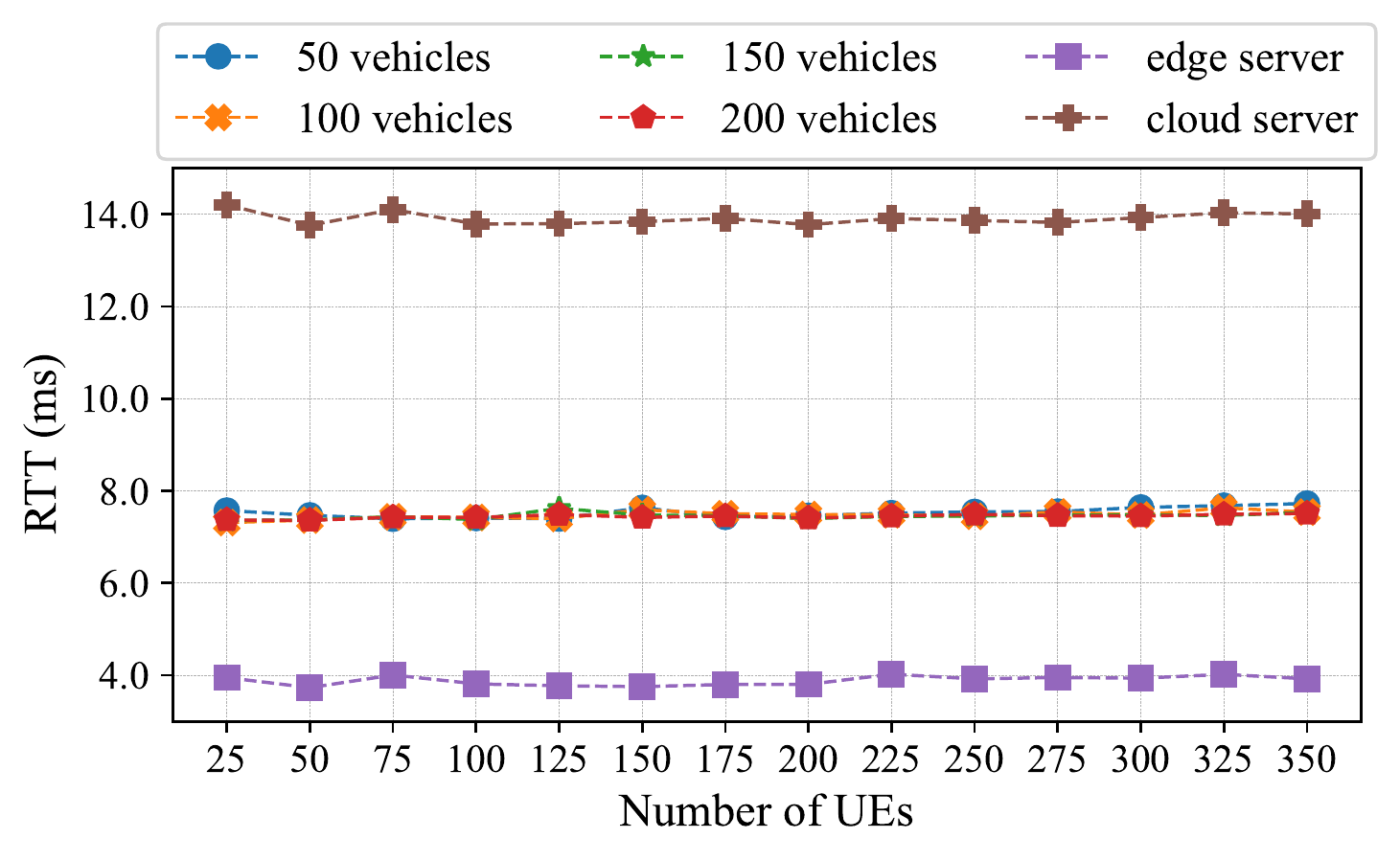}
    \caption{RTT variation with varying number of UE requests in the considered service delivery scenarios. The x-axis denotes the number of UEs requesting an app execution, while the y-axis denotes the average Round-Trip Time between UEs and the \mecapp{}. 
    In the far-edge delivery mode, \mecapp{}s requested by the UEs are deployed onboard the vehicle VI.}
    \label{fig:rtt}
    \vspace{-0.4cm}
\end{figure}

% \input{analysis}
%\begin{table}[htbp]
%\caption{Network settings \af{maybe we should remove the table to have more space. if we want to use it we should add UE parameters as well}}
%\begin{center}
%
%\begin{tabular}{|c|c|}
%\hline
%Parameter name            & Value  \\ \hline
%\#gNB                     & 1      \\ \hline
%gNB numerology index ($\mu$) & 2      \\ \hline
%Carrier frequency         & 2 GHz  \\ \hline
%gNB antenna Gain          & 18dBi  \\ \hline
%gNB tx Power              & 20 dBm \\ \hline
%\end{tabular}
%\end{center}
%\end{table}

\begin{figure*}[ht]
    \begin{subfigure}[b]{0.49\textwidth}
        \centering
        \includegraphics[width=\linewidth, height=0.55\linewidth]{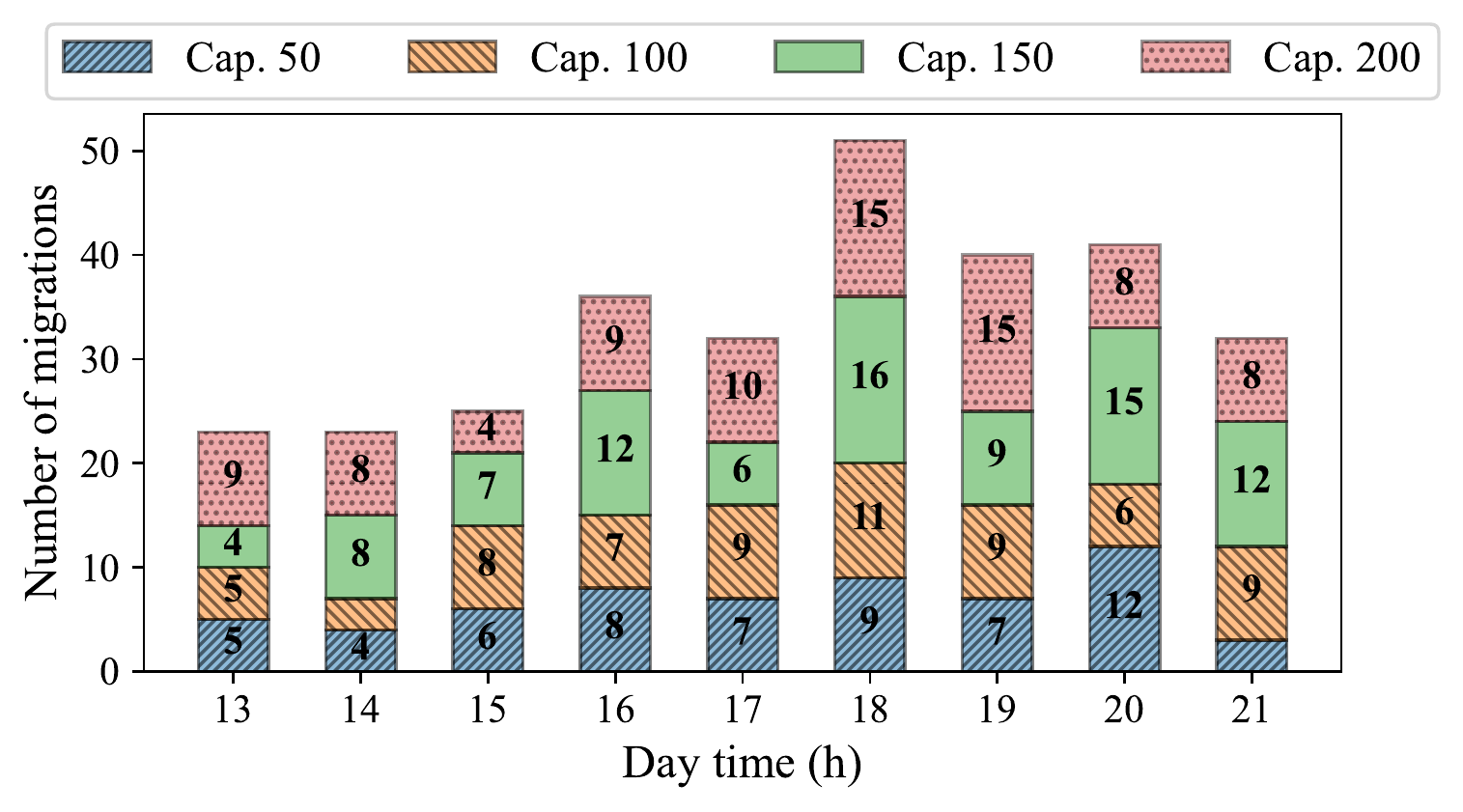}
        \caption{Migration frequency}
        \label{fig:migrations}
    \end{subfigure}
    \begin{subfigure}[b]{0.49\textwidth}
        \centering
        \includegraphics[width=\linewidth, height=0.55\linewidth]{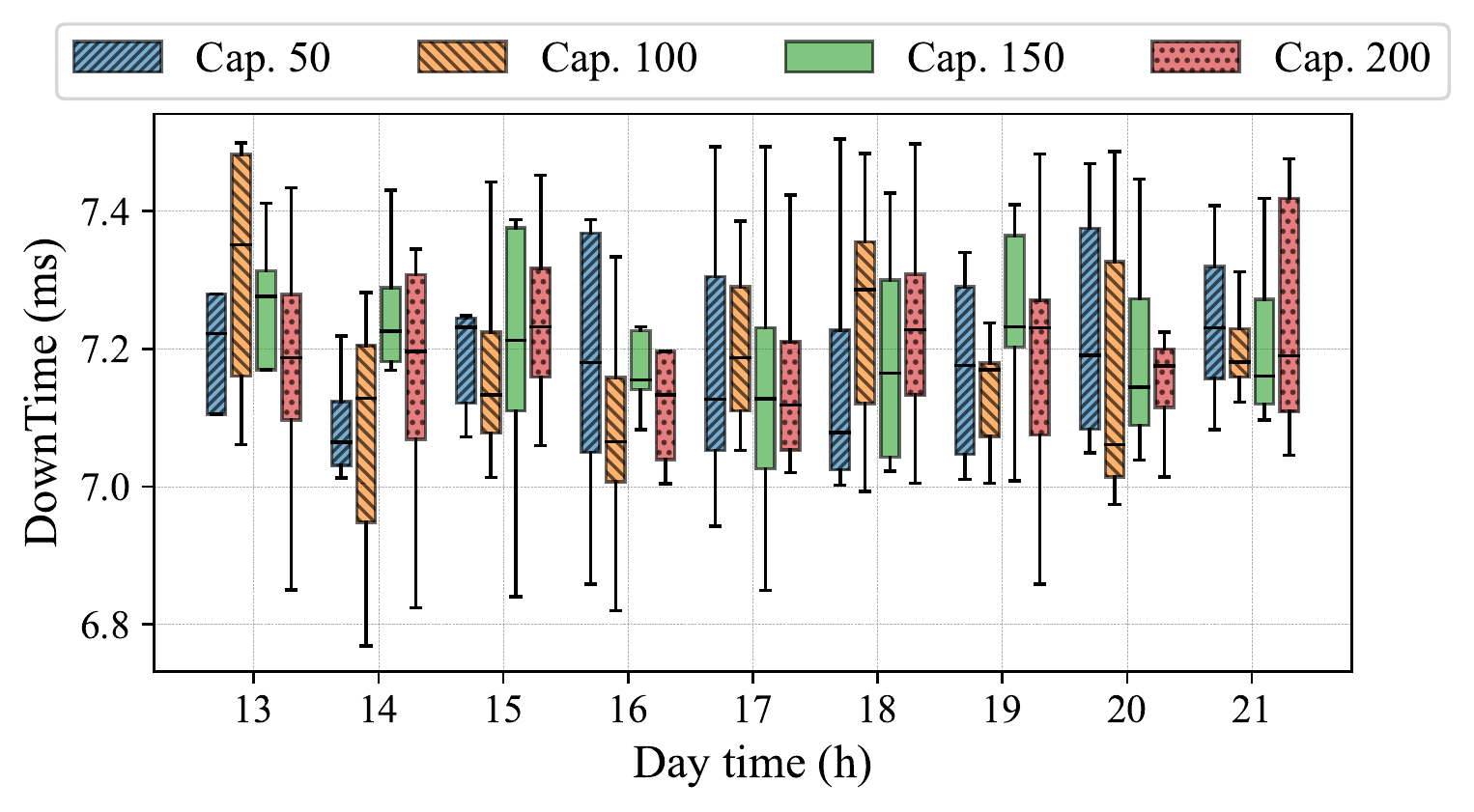}
        \caption{Downtime}
        \label{fig:downtime}
    \end{subfigure}
    \caption{Migration related metrics}
    \label{fig:evaluation}
    \vspace{-0.4cm}
\end{figure*}

\section{Evaluation Results}
We built a simulation model of the proposed ETSI MEC extension in the \omnetpp{} simulation tool, using Simu5G as a communication library, so to demonstrate the viability of our approach in (beyond) 5G scenarios. In the current release, available in \cite{githubrepo}, vehicular nodes are modeled and identified as 5G-enabled User Equipment (UE). The exposed vehicular resources consider the CPU in terms of instructions/second, RAM, and disk space available for the lease. Furthermore, the vehicle model includes a \mec{} \vi{} managing its onboard resources and applications' life-cycle and a module that manages rewards, resource registration, and releasing.

%Armir: 1-3 implementazioni delle design choices. 
% dobbiamo per ora parlare di parking area, perche' non sono stati fatti test in movimento.
At model startup, each \mechost{} identifies its \aoi{} corresponding to an area within the associated gNB coverage. Thus, when a vehicle enters the area, it starts interacting with the \server{}, according to the protocol described in Sec. \ref{sec:sysdes}. As a result, the vehicle joins the \mechost{} resource pool and can be considered by the VIM scheduling logic, which in the current implementation of our model, uses a simple Round Robin algorithm. Note that when vehicles leave the \aoi{}, corresponding app migrations might be triggered: we have decided to leverage only local MEC-H infrastructure as a target platform where to migrate \mecapp{}s.
%This behavior is parametric, and the control logic could be extended to implement more refined mechanisms. 
The rationale behind this choice is to avoid expensive and inefficient domino effects where a vehicle receiving a migrated app leaves the \aoi{}, thus triggering a new migration.

%Our simulation model considers only the local infrastructure as a backup for applications to migrate. In future versions, the model will support other methodologies (e.g., scheduling among other resources in the pool).
%Finally, our model uses the Round Robin algorithm to distribute \mecapp{} on parked cars.

% Il cloud-server non hosta nulla, e' solo usato per effettuare i test. La parte control della core network in Simu5G non e' implementata come componenti distribuite, ma e' centralizzata in un unico componente che ne simula comportamenti basilari. Simu5G permette di agire solo su user-plane
% The adopted 5G network model is the 
In our environment, we adopted a 5G standalone network, encompassing a single \mechost{} and a cloud datacenter to enable comparisons among multiple service delivery modes. The simulation scenario includes a parking area, where vehicles can enter and leave as they want. We currently provide a basic reward scheme, which is always accepted by vehicles partaking the resource acquisition procedure. 
%As a future extension, we envision to generalize the approach by introducing a mechanism supporting pluggable reward schemes.

% cooperative learning scenario?
% We first propose a simulation analysis used to validate the proposed design, comparing delay profiles of different service delivery modes. Next, we propose an innovative scenario, involving vehicular nodes partaking in a cooperative learning scenario, evidencing and profiling the application/state migration feature.  

%All the aforementioned experiments are conducted on 
We have conducted an extensive set of experiments on a Linux VM running \omnetpp{} having 16 CPUs and 64 GB of RAM. It is noteworthy to point out that according to the definition of discrete event simulator, \omnetpp{} does not consider the processing time spent to run the code of any module; hence, all the evaluations concern network-related delays. In the following, we evaluate our model and provide an experimental analysis to assess model performance under dynamic conditions.

\subsection{Model validation}
To validate the effectiveness of our model, we evaluate the network-induced delays in three service delivery schemes, namely at the cloud, edge of the network, and onboard the vehicle. In the cloud scheme, the \mecapp{}s are hosted on a simulated cloud datacenter, thus involving data transmissions spanning the 5G RAN, edge, and core network. The second service delivery scheme embodies delays introduced by running \mecapp{} directly on the \mechost{}, thus those due to 5G RAN and MEC local User Plane Function (UPF) co-located with the gNB \cite{kekki2018mec}. The last scheme involves \mecapp{} running on onboard vehicle resources, i.e., remote host in Fig. \ref{fig:extension_architecture}. In the latter case, network delays are affected by data transmission between the network (i.e., gNBs) and devices (i.e., vehicle running \mecapp{}s and UE requesting their execution). 

%Armir(new): la legenda del grafico mi confondo un'po... perche' evidenziato il 50-100-150-200? Nell'asse X il numero veicoli varie 25-350????  - Spiegato dopo
Figure \ref{fig:rtt} shows the Round Trip Time (RTT) of the above-mentioned schemes. On the x-axis, we considered the number of clients requesting the execution of a \mecapp{} to assess the behavior of the system under different load scenarios. In this experimental setup, we assume that each UE requests precisely the execution of a single \mecapp{} as soon as it enters the gNB coverage. In this scenario, UEs are spawned at the same time, simulating a flash-crowd phenomenon. Furthermore, to avoid any bias in the results, each experiment is repeated 5 times.

% Parked vehicle case - Legato al periodo di prima
%Armir(new): cosa significa "the third scheme has been split?"
The analysis of the third delivery scheme has been conducted by considering different vehicle quantities participating in the resource pool of the \mechost{} namely, 50-100-150-200 vehicles. In this scenario, the \mecapp{} execution triggered by the UEs requests is executed onboard the vehicle. This setting allows us to get further insight into network tolerance and relationships between delays and the number of apps deployed on a single vehicle. The figure shows how the latency times are steady even with more than 500 devices within the coverage of the \mechost{} associated gNB (note that computing-related delays are not considered in the reported simulations).
% , each service delivery mode RTT illustrated remains unchanged when increasing the number of UEs requesting for \mecapp{} execution.}

% Vantaggi parked vehicles
Deploying \mecapp{}s on vehicles brings the advantage of reducing the RTT by around 50\% as opposed to cloud deployment. On the other hand, the far-edge scheme increases the RTT by 3ms if compared with the edge mode, as it exploits fully wireless communications via the 5G network infrastructure. However, it should be noted that Simu5G handles device-to-device (D2D) communications through the gNB base station, thus employing a network-mediated communication model also in D2D scenarios. We expect that the adoption of \emph{sidelink} mode of 5G network would further reduce the RTT, as this enables direct communication between two devices without the participation of a gNB in data transmission and reception.

%\begin{figure}[h]
%    \centering
%    \includegraphics[width=\linewidth]{images/migrations_1.pdf}
%    \caption{Migration}
%    \label{fig:migrations}
%\end{figure}

\subsection{Migration Study}
As mentioned in Sec. \ref{sec:bg}, far-edge nodes may leave the resource pool once their resources have been allocated, resulting in a service disruption. To cope with this problem, the MEC-H initiates a migration procedure, moving running apps from the host leaving the resource pool into another one. However, migration of stateful \mecapp{}s may generate a downtime period in which the involved app becomes unavailable. 
% For instance, a \mecapp{} may involve the decentralized learning, where it coordinated the distribution of a model among several nodes. In such a scenario, when nodes leave the pool the partial results should be transferred to the coordinator \mecapp{}, who started the procedure. 

To measure the associated performance indicators, we build a realistic and innovative scenario involving vehicle volatility and UE activities in a parking area context.
% in the context of a parking area
%where vehicles enter and leave a parking lot, making themselves available for resource sharing, and UEs request for MEC-Application execution. 
To make this scenario as close as possible to reality, we need information regarding vehicle activity (i.e., vehicles joining and leaving an \aoi{}) and UEs activity exploiting a network connection in the surrounding area.
To the best of our knowledge, there are no datasets related to a specific area and collecting information similar to the scenario highlighted above. Thus, in order to recreate such an environment, 
we ground our study and make use of two real-world datasets~\cite{arnhemdataset,bolognadataset}. The first dataset captures vehicle join and leave times in three parking garages in the city of Arnhem, while the other one describes the usage of public WiFi networks in the city of Bologna, recording the number of users joining the WiFi network for each day hour. 

% To link the two dataset and create the described scenario...

To synthetize the aforementioned dynamics, we looked for a relationship between these datasets, analyzing each parking garage and each network available. Therefore, we selected one parking garage context and location, and filtered Bologna's dataset using this information. The location chosen is the \textit{Central garage}, a parking area able to host more than 1000 vehicles and close to the train station of Arnhem. Successively, we extracted data from WiFi networks within the train station of Bologna. After the pre-processing step, we studied the distributions of the occupancy time, i.e., the time a vehicle stays parked, the average number of vehicles entering per hour, and the average number of UEs joining the network per hour. Based on the fitted distribution found for the occupancy time, we decided to approximate it with a normal distribution with mean $\mu=202.80$ minutes and standard deviation $\sigma=135.07$. Other metrics, i.e., the number of entering vehicles and UEs per hour, are better captured by a Poisson distribution as they are independent events that occur randomly within an hour, but with a known average rate. 
% It is noteworthy, that the Poisson distribution describing the vehicle entrance in the parking lot, has been modeled by defining the mean rate based on the ratio between average number of entrance per hour and the park capacity.
%\begin{figure}[h]
%    \centering
%    \includegraphics[width=\linewidth]{images/downtime_1.pdf}
%    \caption{Downtime}
%    \label{fig:downtime}
%\end{figure}

We use the obtained distributions in our simulation model to spawn parked vehicles, hence the requesting UEs. This way, we build a dynamic scenario and simulate vehicles joining and leaving the AoI, and UE requests for \mecapp{} execution. The simulation settings are the same as in the prior scenario, i.e., each vehicle lends its onboard resources, and each UE issues a request to run a \mecapp{}. 
%Armir: sarebbe questa la parte dove si descrive la funct. app? e.g., federated learning?
The employed stateful \mecapp{} logic, provided by Simu5G library, allows the user to define a circular geographic area as a warning zone. The app is responsible for notifying the user whenever it enters and leaves that zone. More complex and meaningful location-based applications could be built, considering innovative scenarios such as decentralized (federated) learning etc.  

%Armir(new): spiegare la "park capacity". generiamo noi sintenticamente queste vehicle vs. abbiamo 4 parcheggi diversi che sn catturati nel dataset? Ce ambiguita. - DONE
Figure~\ref{fig:migrations} illustrates the number of migrations generated adopting the distributions described above. We considered four arbitrary parking lot capacities (i.e., 50, 100, 150, and 200) used to scale quantities generated by the Poisson distribution of the original dataset describing vehicles' entrances in the parking lot. As \mecapp{} s are equally distributed among the parked cars using Round Robin algorithm, lower park capacities may correspond to a greater number of migrations, for instance, during day time at 15:00 and 20:00. Note that the number of migrations also depends on the user activities and the amount of parking cars in that hour of the day, both generated through probability distributions. Thus, the relocations occurred may vary according to the number of users requesting for app execution and available car resources, which are not illustrated in this paper. Furthermore, we analyzed the interval from 13:00 to 21:00, as it corresponds to day hours with more network activity. 

Finally, the downtime has been measured by considering the elapsed time between the shutdown of the app running on the leaving host and the end-user receiving the new \mecapp{} location.
% the resuming of the migrated app on \mechost{}. 
Generally speaking, the downtime may be affected by latency between the two involved entities, i.e., remote and local \vi{}, and bandwidth~\cite{7399400}. In our experiments, the former includes the delays due to \mechost{} distance from the gNB and 5G radio delays, while the latter does not affect the service interruption time, as the state sent by \mecapp s is smaller than 30B.
%The scenario allows us to evaluate the \mec{}-App downtime in real-world conditions. 
%Armir: espandere di piu' il trend grafico! Bisogna essere piu' argomentativi... dando cifre, delta etc.
% Il trend del grafico e' stabile come gia' detto nelle righe qui sotto. Quello che ne emerge da questa analisi e' specificato enlle ultime due righe: il valore del migration time rimane costante nonostante entrambi numero di mitrazioni e carico della rete aumentino.
Hence, as shown in Figure~\ref{fig:downtime}, the downtime remains stable around 7 ms.%, with some peaks near 4 ms reached with lower park capacities.
Overall, despite the number of \mecapp{} relocations (Fig.\ref{fig:migrations}) and the network activity increase, the downtime remains stable when migrating \mecapp{}s from one host to another. %Hence, the results demonstrate the reliability of our solution in a dynamic and high-loaded real-world scenario.

%Armir: parte sotto e' piu' da conclusion. Commentiamo per ora.
%The survey provided in this section allowed us to replicate in our simulation model dynamic real-world conditions, where vehicles join and leave \mechost{} remote resource pool. Consequently, we managed to give a realistic evaluation of our simulation model downtime under these conditions. Furthermore, the presented survey represents the first effort in providing the community vehicular cloud behaviors in a real-world scenario. 
% \af{may be change the last sentence}

% In this environment, each UE generates a \mec{} compliant request for \mecapp{} execution. Thus, 

% \af{note that simulator does not simulate computational delays}

% \af{give some details about the mec app used}

% \af{Cite this \cite{6143927}}

\section{Conclusions and Future Work}
In this paper, we propose a novel architecture extending the ETSI MEC standard to enable the dynamic negotiation and acquisition of far-edge resources at \mechost{} level. The presented scheme allows device owners to access a reward system and addresses resource volatility problems as devices join and leave the \mechost{} resource pool. To demonstrate the viability of our approach, we built a simulation model that extends ETSI \mec{} towards scenarios with dynamic availability of vehicular resources. The model is validated with three service delivery schemes, thus demonstrating how our approach introduces good performance in \mec{}-enabled environments. Finally, we also proved its robustness in real-world dynamic conditions.

Although the nature of our architecture considers both physical and mobile nodes, in this discussion we have limited the simulation-based evaluation to parked vehicles and local migrations within the \mechost{}. As future work, we envision expanding our simulation model to support resources provided by stationary and moving vehicles in more complex scenarios, e.g. smart-city ones. Consequently, we plan to generalize the reward mechanism to support pluggable reward schemes, e.g., relying on user behaviors. Moreover, we plan to extend the support for pluggable scheduling modules so to study the effect of scheduling strategies that distribute and migrate \mec{} applications among far-edge nodes according to device parameters (e.g., energy consumption) and app requirements (e.g., max latency).

\bibliographystyle{IEEEtran}
\bibliography{references}
\balance

\end{document}